\documentclass[prd,eqsecnum,aps,showpacs]{revtex4-1}

\usepackage{epsfig}
\usepackage{latexsym}
\usepackage{amssymb}
\usepackage{amsmath}
\usepackage{epsfig}
\usepackage{color}
\usepackage{textcomp}

\newcommand{\beq}{\begin{equation}}
\newcommand{\eeq}{\end{equation}}
\newcommand{\bea}{\begin{eqnarray}}
\newcommand{\eea}{\end{eqnarray}}

\begin{document}
\title{Inflationary field excursion in broad classes of scalar field models}
\author{Argha Banerjee}
\email {baner124@umn.edu}
\affiliation{ Department of Physics, Presidency University,
86/1 College Street,
Kolkata - 700 073, India \\ School of Physics and Astronomy, University of Minnesota, Minneapolis, MN 55455 USA}
\author{Ratna Koley}
\email {ratna.physics@presiuniv.ac.in} 
\affiliation{ Department of Physics, Presidency University,
86/1 College Street,
Kolkata - 700 073, India}

\begin{abstract}

In single field slow roll inflation models the height and slope of the potential  are to satisfy certain conditions, 
to match with observations. This in turn translates into bounds on the number of e-foldings and the excursion of the scalar field during 
inflation. In this work we consider broad classes of inflationary models to study how much the field excursion starting from horizon 
exit to the end of inflation, $\Delta \phi $, 
can vary for the set of inflationary parameters given by Planck. We also derive an upper bound on the number of e-foldings between
the horizon exit of a cosmologically interesting mode and the end of inflation.  We comment on  the possibility of 
having super-Planckian and sub-Planckian field excursions within the framework of single field slow roll inflation.
 \end{abstract}
\maketitle

\section{Introduction}

The standard big bang cosmology has been proved to be successful in explaining the observed evolution of the universe, albeit with 
some extremely fine tuned initial conditions. The era of cosmological inflation \cite{AG,Al} was introduced to take care of such 
initial conditions, and it provides a very nice proposal for the solution to the horizon problem, the flatness problem and a very good 
explanation for the 
nonexistence of unwanted relics. The most salient feature of inflation is the quantum fluctuations that render seeds for the 
large scale structure, together with a possible gravitational wave contribution, for the cosmic microwave background (CMB)
anisotropy \cite{Ba1, LL2}.
In its simplest form, inflation is best realized by means of a  minimally coupled scalar field in the framework of Einstein gravity. Recent 
CMB data by Planck 2015 \cite{Pl} indicate that the power spectrum of density perturbations of the scalar field 
is nearly scale invariant, which is apparent from the value of the scalar  spectral index, $n_s  = 0.968 \pm 0.006$. Planck has also taken the 
cosmologists by surprise by predicting an almost Gaussian nature of the power spectrum and putting only an upper bound,  $ r <  0.11$,  
on the amplitude of primordial gravity waves by considering a tensor amplitude as a one-parameter extension to the $\Lambda CDM$ model.  
An even tighter bound, $r < 0.09$ has been obtained by combining Planck with BICEP2/Keck likelihoods \cite{Pl}.  
The BICEP2/Keck array VI reports even more tighter bound on the tensor to scalar ratio, $r<0.07$ when the above mentioned 
constraints from Planck analysis of CMB temperature are combined with BAO (Baryon Acoustic Oscillations) and other data \cite{Ke}. The height and the slope of the inflaton 
potential must maintain a delicate balance for the 
compatibility with observations. This in turn translates into the excursion of the scalar field during the horizon exit to the end of inflation. 
In this work we want to address the question that how much the field excursion can vary for a given set of inflationary observables. 

The magnitude of the stochastic background gravitational waves,  for single field slow roll inflation, is related to the energy scale of inflation and more importantly, 
it is linked to the inflaton excursion. In the standard single field slow roll inflationary scenario, according to the Lyth bound \cite{lyth}
a  sizable detection of tensors would mean a super-Planckian excursion of the inflaton via the constraint 
$r \lesssim 0.01 (\Delta \phi /M_{Pl})^2$ where
$M_{Pl}$ is the reduced Planck mass. This is definitely interesting both from the model building standpoint and from an observational 
one. The original Lyth bound can be evaded if one considers non-slow roll inflation {\cite{lythbau}} or simply considers 
extra sources for density perturbations {\cite{lythlinde}} or has additional light degrees of freedom contributing to the production of 
perturbations {\cite{lythtaka}}. Other theoretical bounds on the tensor fraction as a generalization of the original Lyth bound has been 
discussed in {\cite{lythb}}. The amount of inflation between the horizon exit of a cosmologically relevant mode  
 and the end of inflation is given by 
\beq
N = N_* - N_f = \int^{a_f}_{a_*} d \ln a
\eeq
where the subscript $*$ means that the quantities are evaluated at horizon exit  
 and $f$ means that at the end of inflation respectively, $a$ is the scale factor. To address the horizon problem and subsequently all others, it is necessary to have at least 50 e-foldings in this period in the conventional inflationary scenario. So observationally we can only fix a lower  limit for $N$ but  
 there is no compelling evidence of any upper limit on the total
  amount of inflation. In fact  it may be extended a long way further into the past than the present horizon size. By using the phase space analysis in foliating FRW (Friedmann Robertson Walker) universe this possibility has been explored in {\cite{carroll}}. 
  
Our aim, in this work, is to determine how much the field excursion $\Delta \phi= | \phi_* - \phi_f |$ can vary
 given an inflaton reproducing observed cosmological parameters such as $n_s$, $\alpha_s$ and $r$, . To this end we consider the classification of single field slow roll 
inflationary models as demonstrated in \cite{rb, roest}. All those models whose slow roll parameters scale with 1/N or a higher power can be
classified into two broad categories characterised by a single parameter  $\Delta \phi$, the field excursion. By expressing the inflationary 
observables in terms $N$ one can also group the models of inflation into broad classes like constant, perturbative, non-perturbative and 
logarithmic \cite{rb, roest}. This large-$N$ formalism is a more effective way of studying the inflationary models instead of doing the case-by-case 
analysis. 

Now as the benchmark we choose a model of inflation with a strong field theoretical background, 
which passes successfully through the observational constraints set by recent CMB data.  There can be many viable phenomenological 
models which fit well with observations.  In this article the choice for the benchmark model has been made by giving stress on high energy theoretical 
background. We select a model which arises in the context of type IIB string theory via Calabi-Yau flux compactification. One such example is where 
one of the K\"{a}hler  moduli playing as inflaton when internal spaces are weighted projective spaces in  type IIB string theories \cite{km}. 
 The version  with  the  canonically normalized inflaton field known as  K\"{a}hler Moduli  II (KM II) inflaton \cite{km3} has been chosen as 
 benchmark in our case. Most importantly this model can be understood in the context of supergravity, viewed as an effective theory.
It has been the general practice  earlier \cite{belli} to choose the 
 chaotic inflationary scenario \cite{chaotic} as benchmark. However the minimal chaotic models  are almost ruled out after Planck 2015 for 
 not satisfying the bound on stochastic gravity wave amplitude (for the chaotic model $r > 0.09$).  In addition to
that the BICEP2 results giving large value of $r$ have also been discredited, therefore one cannot be sure about the
benchmark status of the chaotic model. We are interested to explore the effect of $n_s$ and
$r$ on the field excursion by considering the observational bounds set by the recent Planck data. Now the
KM II model of inflation gives very low value of $r$, thus giving a sub-Planckian field excursion. 
Given the fact that the benchmark model passes all observational tests we find it to be a pertinent question to ask whether the field 
excursion of inflation should be in the same range of the benchmark or not?  To this end we would like to explore the effect of $n_s$ and 
  $r$ on the field excursion by considering the observational bounds set by Planck \cite{Pl}.

  \section{Asymptotic Hubble flow functions in KM II inflation :} 

We start by recalling the basics of the KM II model \cite{km, km3} of inflation and finding the Hubble flow functions in the large $N$ formalism. The potential is given by 
\beq
V (\phi) = V_0 \left[1-\alpha\left(\frac{\phi}{M_{Pl}}\right)^{4/3}\exp\left({-\beta\left(\frac{\phi}{M_{Pl}}\right)^{4/3}}
\right)\right]
\eeq 
Making use of the typical orders of magnitude one can write the parameters $\alpha$ and $\beta$ as 
\beq
\alpha= \mathcal{O}\left({\cal V}^{5/3}\right), ~~~~~~~~  \beta=\mathcal{O}\left({\cal V}^{2/3}\right)
\eeq
where the quantity $\cal V$ represents the Calabi-Yau volume. The potential starts from a maximum, $V=V_0$ at $\phi = 0$, then reaches the minimum at $\frac{\phi}{M_{Pl}} = \beta^{-3/4}$ and finally asymptotes to
$V=V_0$ for $\frac{\phi}{M_{Pl}}$ approaching $\infty$. Maintaining the consistency with reheating, the slow-roll predictions 
for the KM II model can be achieved for ${\cal V }\in [10^5,10^7]$ and thus the parameters $\alpha$ and $\beta$ 
can have values in the range $ \alpha \in [2.15 \times 10^8, 4.64 \times 10^{11}]$ and $\beta \in [2.15 \times 10^3, 4.64 
\times 10^4].$ It can be shown that the Hubble slow roll predictions do not depend significantly on the vales of $\alpha$ and $\beta$ \cite{martin}. We 
now intend to find out the Hubble flow functions $\epsilon_n$ defined as \cite{largeN} 
\beq
\label{enplus1}
\epsilon_{n+1}= \frac{d}{dN} \log|\epsilon_n|,~~~~ n \geq 0
\eeq
 
 for large $N$ in case of KM II model of inflation. The above functions basically play the role of slow roll parameters in standard 
 formulation in terms of $\phi$. Here $\epsilon_0$ is nothing but the Hubble parameter and the range of $N$ runs starting from horizon exit to the end of inflation. As $N$ depends on derivatives of $V(\phi)$ it is apparent from (\ref{enplus1}), that the successive Hubble flow functions are  related to the derivatives of the potential $V(\phi)$. Consequently  the slow roll parameters can also be expressed in terms of the Hubble
 flow parameters varying as $1/N^p$ for some values $p$ at leading order in the limit of large $N$. This will become evident from the following calculations. Further at first order in $\epsilon_n$ one
 can represent the CMB observables of inflation as 
\bea
\label{nr}
n_s&=&1 -2\epsilon_1 + \epsilon_2 \\ r &=& 16\epsilon_1.
\eea
  To set up a connection with the observables one needs to calculate these quantities at the time of horizon crossing of a cosmologically relevant scale. It has been noticed by Lyth 
\cite{lyth} that the tensor to scalar ratio of temperature fluctuations i.e. the first slow roll parameter can be related to the field excursion via the relation 
 \beq
 \label{phi}
\frac{1}{M_{Pl}} \frac{d\phi}{ dN} \sim \sqrt{\frac{r(N)}{8}} = \sqrt{2\epsilon_1}.
\eeq

Assuming $r(N)$ to be invariant throughout the phase of inflation it can be shown that \cite{lyth, lythb} the field excursion is
\beq
\Delta\phi\approx \left(\frac{r}{0.002}\right)^{1/2}\left(\frac{N_*}{58}\right)M_{Pl}
\label{lythbound}
\eeq
 where ${N_*}$ is set to 58 which falls within the range of  ${N_*}$ allowed by Planck \cite{Pl} pivot scale. 
However this particular value has been chosen arbitrarily within the permissible range. It is apparent 
from the above equation ({\ref{phi}}) that for $r < 0.002$ we have $\Delta\phi < M_{Pl}$ leading to sub-Planckian field excursion 
while for $r > 0.002$ we get super-Planckian field excursion. As a result one can distinguish the inflationary models in terms of 
the field excursion variable. 

Now we are all set to calculate the Hubble flow functions for the KM II model given in (\ref{enplus1}). From the observational point 
of view one needs $N$ to be large and thus these parameters are of singular importance for the rest of the analysis as we will see 
that there are large classes of models that agree on large $N$ limit. The first order Hubble flow function is given by
\beq
\label{kmsr1}
\epsilon_1=\frac{b}{2N^2\sqrt{\ln N}}
\eeq
where $b=\frac{9}{16} \frac{1}{\beta^{3/2}}$ and the second Hubble flow function is as follows
\beq
\label{kmsr2}
\epsilon_2= - \frac{2}{N}.
\eeq
The basic features of the inflationary model under consideration  have been encoded by the above functions at the leading order of $N$. 
Subsequent correction terms have very insignificant role to play with the observational parameters. Let us consider that from now on the 
quantities of the benchmark model will be denoted by an overhead bar to differentiate them from the other classes of inflation and choose 
to work with $M_{Pl} = 1$.

We take three values of ${\cal {V}} = 10^5, 10^6$ and $ 10^7$ for our analysis, leading to the values of  
$b = 5.63\times 10^{-6}$, $5.63\times10^{-7}$ and $ 5.63\times 10^{-8}$ respectively.
As most of the inflation takes place at large values of $N$ we can consider $N_f$ to be negligibly small and thus $\bar N_*\approx 58$
is justified. This particular choice for the number of e-folds remaining after the exit of horizon to the end of inflation is 
in agreement with the Planck pivot scale. Other allowed values of $\bar N_*$ may be chosen but that will only enable us to infer similar 
output for the analysis. Let us now define a quantity as follows 
\beq
\label{e1*}
\epsilon_{1*}=\frac{b}{2\bar N_*^2\sqrt{\ln \bar N_*}}
\eeq
which is the value of the first Hubble flow parameter at horizon crossing and $\bar N_*$ is the no. of e-folds at that point of time.  
As the benchmark matches very well with observational parameters, we set our aim 
of study to learn how much these predictions are compatible with universality classes of inflationary models which agrees in the large $N$ 
limit. We are also curious to know what happens to the field excursion variable $\Delta \phi $ for the broad classes of models mentioned 
earlier in comparison with the KM II model.

\section{Comparison of field excursion in different classes of inflaton}

We now intend to look how the field excursion of the inflationary 
models vary for a given set of values of the CMB observables $n_s$ and $r$. It will be interesting to explore whether the 
field excursion $\Delta\phi$ and the number of e-folds $N$ remain the same or change. The large $N$ behaviour of wide 
classes of inflationary models have been discussed rigorously in \cite{roest,rb} by finding the dependence on $N$ of 
the Hubble flow parameters. At leading order the $1/N^p$ behaviour for the slow roll parameter $\epsilon$ is considered as 
the {\em{perturbative}} class. In addition to that there are {\em{constant, non-perturbative and logarithmic classes}} \cite{roest,rb}. 
For these three classes we will find the leading order contribution of the first and second Hubble flow parameters and 
equating those to the respective 
values for the KM II model we will compare the field excursion for a given set of spectral tilt $n_s$ and tensor-to-scalar ratio $r$. \\

\subsection{Perturbative Class}
An attractive feature of the {\em{perturbative}} class of models  is that
the $1/N$ term provides a natural explanation for the percent variation 
from the scale invariance of the CMB power spectrum.  Chaotic, hilltop, inverse hilltop, Whitt potentials
are typical examples of this particular class. The first two Hubble flow parameters 
of the perturbative class are given by
\bea
\label{pereps}
\epsilon_1=\frac{\mu}{N^p} \nonumber \\ \epsilon_2= - \frac{p}{N}
\eea
where $\mu$ and $p~ (\geq 1)$ are the parameters, for different values of which one gets different 
 models within this class. Now by imposing the requirement that the above Hubble flow parameter values should fall in the same range 
 as that of the benchmark  model as given in Eq. (\ref{e1*}), i.e. the same scalar spectral index and tensor-to-scalar ratio will be 
 produced by the perturbative class of models as that of the benchmark, we obtain the following relationship to be followed by the model
 parameters. Let us consider first the parameter $\mu$ which should follow the restriction given below to reconcile with the above 
 mentioned demand.  
\beq
\label{p1}
\mu=\frac{b N^p_*}{2 \bar N^2_* \sqrt{\ln\bar N_*}}
\eeq
where 
\beq
\label{p2}
N_*=\frac{p \bar N_*}{2}
\eeq
Terms with an over bar correspond to the values associated with the benchmark model. 
Now substituting equation (\ref{p2}) into the equation (\ref{p1})  we obtain 
\bea
\label{p3}
\mu&=&\frac{b}{2}\left(\frac{p\bar N_*}{2}\right)^p \times \frac{1}{\bar N_*^2\sqrt{\ln \bar N_*}} \nonumber \\
&=& \epsilon_{1*} \left(\frac{p\bar N_*}{2}\right)^p
\eea

 Equation (\ref{p3}) explicitly indicates how the parameters should be finely adjusted to guarantee the correct prediction 
of observational parameters coming from 
up-to-date CMB observations in several models in the perturbative class. A careful 
investigation on how the parameter $\mu$ behaves for wide range of $p$, reveals  
that one can get the same values of $b$ for diverse values of $\mu$ and $p$.
This in turn says, as we fix the values of the slow roll parameters of the perturbative model with that of the KM II model, 
the slow roll parameters of the perturbative model and thus subsequently the values of the spectral index $n_s$ and 
tensor-to-scalar ratio $r$ are fixed while $\mu$ and $p$ change.


Let us  explore the number of e-folds $N$ from horizon exit to end of inflation and the field excursion 
$\Delta \phi$ for large classes of perturbative models characterised by different values of $p$. It is apparent from the definition in 
Eq.(\ref{enplus1}), the end of inflation can be associated to the first Hubble flow parameter, $\epsilon_1 = 1$. 
 The number of e-folds, 
$N_f$, at the end of inflation can thus 
be determined from the above mentioned condition. From  Eq. (\ref{pereps}) we get
\beq 
\label{pne}
N_f=\mu^{1/p}
\eeq
Therefore the number of e-folds $N$ for the perturbative class in terms of $p$ and benchmark model parameters is obtained 
by using the above value of $N_f$ and Eqs. (\ref{p2}) and (\ref{p3}) as given below
 \bea
 \label{pn}
 N&=& \frac{p\bar N_*}{2}\left[1- \epsilon_{1*}^{1/p}\right]
 \eea
 \begin{figure}[h] \begin{center}
\includegraphics[scale=0.7]{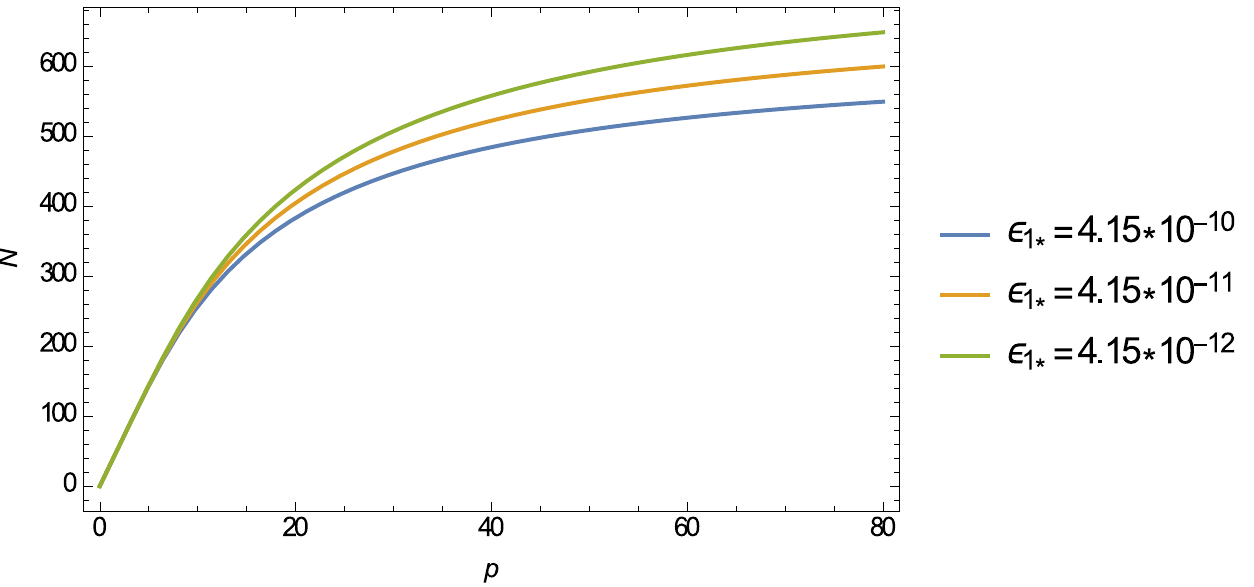} 
\caption{ $N$ increases linearly for low values of $p$. Three values of $\epsilon_{1*}$ are used for three values 
of $\beta = 2.15\times10^3, 1.00 \times10^4,4.64\times10^5$.}
\label{pertn1} \end{center} \end{figure}
 For a given $b$, the value of $\epsilon_{1*}$ being very small it becomes apparent from the above relation that $N$ increases linearly 
 with $p$ for low values which is evident from Fig. ({\ref{pertn1}). One may find it interesting to allow $N$ to vary for a large range which 
 may be dependent on the post inflationary  physics of the model. Curiously, we have noted (in Fig.{\ref{pertn2}) that a maximum limit on the 
 value of $N$ is reached asymptotically with $p$ and this seems to be a generic feature for this class of models. The consequences
 of this finding will be explored further by studying the field excursion.
 \begin{figure}[h] \begin{center}
\includegraphics[scale=0.7]{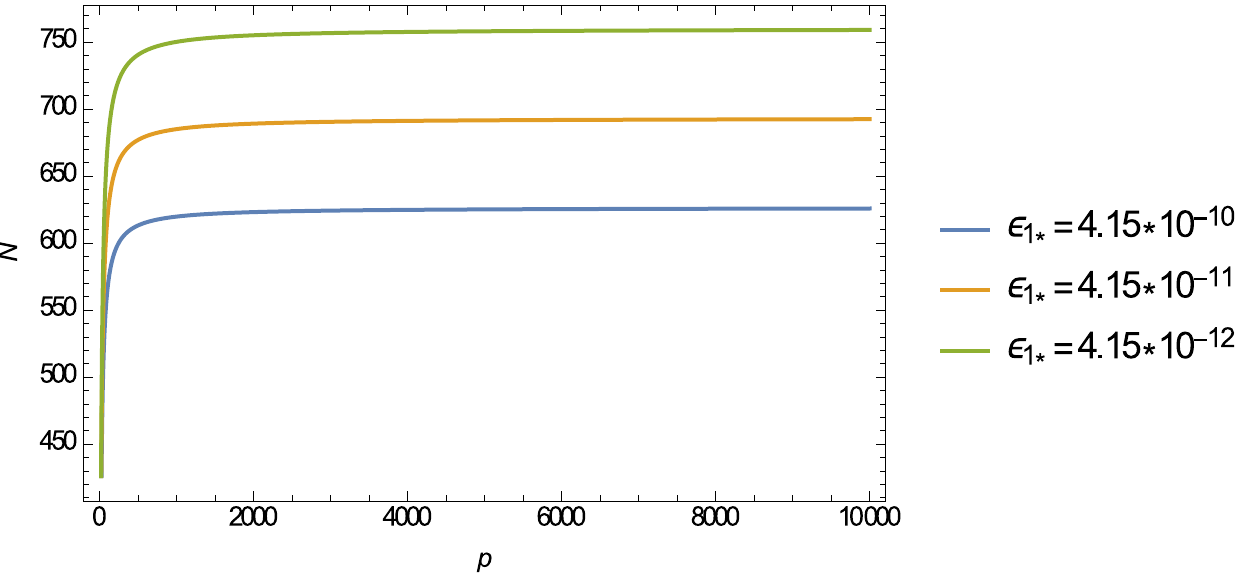} \caption{$N$ approaches a maximum value for high values of $p$.}
\label{pertn2} \end{center} \end{figure} 
Using the definition given in Eq. (\ref{phi}) we get the excursion of inflaton as follows  
\beq
\Delta \phi=\frac{2\sqrt{2\beta}}{2-p}\left(N_*^{1-p/2} - N_f^{1-p/2}\right)
\eeq
  for the perturbative class. Putting in the values of $N_f$ and $N_*$  the above expression reduces to the elegant form 
   \beq
 \label{delphip}
 \Delta\phi=\frac{\sqrt{2}}{2-p}(p {\bar{N}}_*)\epsilon_{1*}^{1/2}\left[1-\epsilon_{1*}^{\frac{2-p}{2p}}\right].
 \eeq
 Let us depict the results graphically by the Fig. (\ref{perdel1}) and (\ref{perdel2}) 
 which show the  variation of  $\Delta\phi$ with respect to $p$. Interestingly, $\Delta\phi$ starts out as 
 sub-Planckian ($\Delta\phi < M_{Pl}$) for small values of $p$ before it becomes equal to 1 ( we choose to work 
 with $M_{Pl} = 1$) at a certain value of $p$. Beyond that a continuous increase is seen in $\Delta\phi$ as $p$ 
 increases finally saturating at high values of $p$. This also shows an upper bound on the value of the field 
 excursion similar to what is found in the number of e-folds.
 \begin{figure}[b] \begin{center}
\includegraphics[scale=0.7]{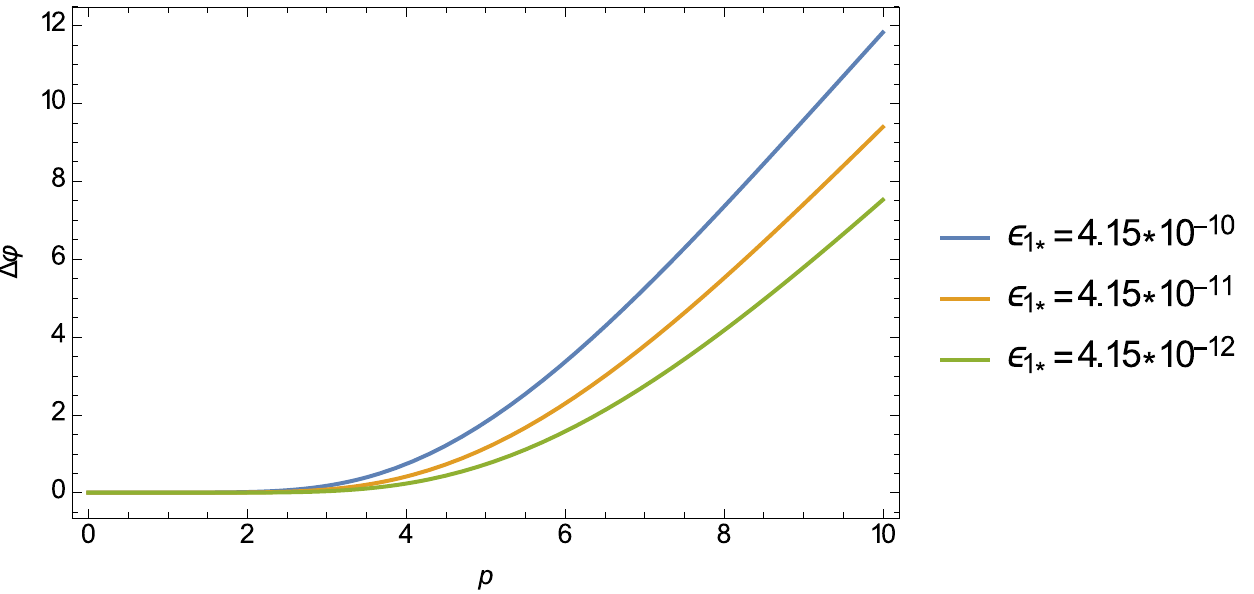} \caption{$\Delta\phi$ increases linearly for low values of $p$ 
and is sub-Planckian up to a certain value of $p$ beyond which becomes super-Planckian. Three values of 
$\epsilon_{1*}$ is used to span the entire range of $\beta$ by choosing $\beta = 2.15 \times 10^3, 1.00 
\times 10^4, 4.64 \times 10^5$.}
\label{perdel1} \end{center} \end{figure}
 \begin{figure}[h] \begin{center}
\includegraphics[scale=0.7]{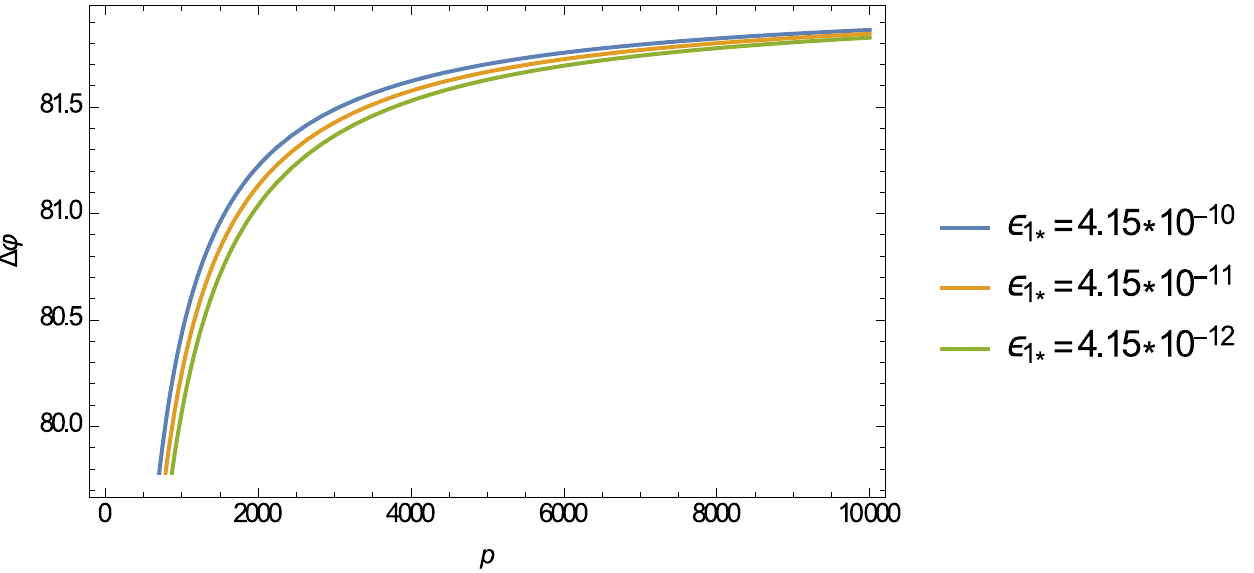} \caption{$\Delta\phi$ becomes super-Planckian for $ p>4$ with the given 
parameter choice and  approaches a maximum value for high values of $p$. Here we have taken 
$\beta = 2.15 \times 10^3, 1.00 \times 10^4,  4.64 \times 10^5$ which lead to the values of $\epsilon_{1*}$ shown above.}
\label{perdel2} \end{center} \end{figure}

 One can easily find the maximum values of the number of e-folds and the field excursion by looking at the limiting 
 tendencies as $p$ goes to infinity. Let us discuss one example by choosing a typical value of  
 $\epsilon_{1*}\approx 10^{-10}$. We find that 
 \bea
 \label{nmaxp}
 N_{max}=\lim\limits_{p\rightarrow\infty} N&=&\lim\limits_{p\rightarrow\infty} \frac{p\bar N_*}{2}\left[1- \epsilon_{1*}^{1/p}\right] \nonumber
 \\ &=& -\frac{\bar N_*}{2}\ln{\epsilon_{1*}}=625.99
 \eea
 \bea
 \label{delphimaxp}
 \Delta\phi_{max}=\lim\limits_{p\rightarrow\infty} \Delta\phi&=& \lim\limits_{p\rightarrow\infty} \frac{\sqrt{2}}{1-2/p}\bar N_* 
\epsilon_{1*}\left[1-\epsilon_{1*}^{1-p/2} \right] \nonumber \\ &=&\sqrt{2}\bar N_*\left[1-\epsilon_{1*}^{1/2}\right]=82.02
 \eea
However such large values of $N$ are not necessarily realistic, from a theoretical view point it is interesting to explore such large ranges. 
Considering the variation of $\epsilon$ with respect to $N$ one can show that it is impossible to keep $\epsilon$ constant for a large
range of e-foldings. As a result there appears an upper bound on $N$ which translates into a limit on field excursion.  

 People have been curious for long about how deep the inflation can be in specific classes of models. Lyth bound gives a 
 guideline for the minimal single field slow roll scenario. Depending on whether one considers $\epsilon$ to vary or not 
 during the horizon exit to the end of inflation more stringent constraints of Lyth bound can be imposed  \cite{roest1}.  In this analysis 
 we retain the considerations originally used to define the field excursion. From the  main results obtained in the perturbative 
 class we see that both the field excursion and the number of e-folds increase with increase in $p$ even as  $n_s$ and $r$ remain the same. 
 Most remarkably an upper limit on both $\Delta\phi$ and $N$ has been achieved asymptotically. A careful inspection points towards a 
 degeneracy in the 
 field excursion with different values of $N$ for the  models predicting same values of observational parameters. We have also explored 
 the possibility of having sub-Planckian and super-Planckian field excursion. This is very interesting both from theoretical and observational 
 point of view. We only have a bound on the tensor-to-scalar ratio from the observations till today. A definite detection of $r$ will 
 definitely solve the above puzzle also an independent detection of $r$ and $\epsilon_1$ is necessary to prove the validity of Lyth bound. Note 
 that the  bound given in Eq. (\ref{lythbound}) implies that for sub-Planckian inflaton excursion and thus consistent field theory description 
 $r$ should be less than $ 0.002$, implying that it was beyond the reach of Planck but within reach of future missions like various ground-based 
 experiments (AdvACT, CLASS, Keck/BICEP3, Simons Array, SPT-3G), balloons (EBEX 10k and Spider) and satellites (CMBPol, COrE and LiteBIRD)
 \cite{kogut, Creminelli}.\\

 
 \subsection{Non-perturbative class}
 
The next class that we would like to consider is the non-perturbative models of inflation \cite{roest, rb} characterised 
by the Hubble flow parameters which are non-perturbative around $1/N \rightarrow 0$. In this case
\bea
  \label{e1no}
  \epsilon_1=\exp{(-2cN)} \nonumber \\ \epsilon_2= -2c
  \eea
  where $c$ is a constant. Equating this with the same parameters of the benchmark model  we obtain 
  \beq
  \label{n*no}
  \frac{b}{2\bar N_*^2\sqrt{\ln \bar N_*}}= \exp{[-2cN_*]}
  \eeq
  leading to an expression for the constant $ c= {1}/{\bar N_*}$. We next consider the number of e-folds $N$ between horizon exit and
  end of inflation. The number of e folds at the end of inflation, $N_f$ is obtained by the fact that the first Hubble flow parameter is equal
  to 1 when inflation ends. 
Thus for the non-perturbative case we have $N  \approx N_*$. Using equation (\ref{n*no}) we can calculate the number e-foldings 
remaining at the point of horizon exit as
 \beq
 \label{n**no}
 N_*= -\frac{\bar N_*}{2} \ln {\epsilon_{1*}}
 \eeq
This in turn gives the number of e-foldings in the non pertubative class from horizon exit 
to end of inflation as  $N = -\frac{\bar N_*}{2} \ln {\epsilon_{1*}}$. 
We get a startling result for the number of e-folds. The number of e-folds $N$ is the same as the form for the 
maximum number of e-folds for the perturbative class (Eq. \ref{nmaxp}). 
Apparently the $N$ of the non-perturbative class hits the maximum limit of the number of e-folds for the perturbative class.
 
 The field excursion $\Delta\phi$ as obtained using Eqs. (\ref{phi}) and (\ref{e1no}) has the following form 
 \beq
 \Delta\phi=-\sqrt{2}\bar N_*\left[\exp{(- c N_f)} - \exp{(-cN_*)}\right].
 \eeq
 Inserting the value of $N_f$ and using the Eq. (\ref{n**no}) we get
 \beq 
 \Delta\phi=\sqrt{2}\bar N_*\left[1-\sqrt{\epsilon_{1*}}\right]
 \eeq
which is same as that for the maximum limit of $\Delta\phi$ for the perturbative class (Eq. \ref{delphimaxp}).  
Most significantly $N$ and $\Delta\phi$ in the non-perturbative class is similar to that 
in the large $p$ limit of the perturbative class. Furthermore  there is not much variation over the 
different parameters, instead there is one particular value of $\Delta\phi$ and $N$ each for different values of $\epsilon_{1*}$ 
corresponding to the benchmark model KM II. Curiously, for non-perturbative class we get super-Planckian field excursion only. 
This is actually a very strong constraint because first of all it is difficult if not impossible to have one inflationary theory where 
we have a good control over a Planckian field range. It again establishes the necessity for independent detection of first Hubble 
flow function and $r$ that will tell us about the existence of Lyth bound.
 
\subsection{Logarithmic class}

The other class of model that we intend to explore is the logarithmic one \cite{rb} in which the Hubble flow parameters are 
\beq
\epsilon_1=\kappa\frac{\ln^q N}{N^p}
\eeq
\beq
\label{le2}
\epsilon_2=-\frac{p}{N}+\frac{q}{N\ln N}
\eeq
where $p$ and the power coefficient $q$ are model specific parameters different values of which lead to different models. 
We have retained logarithmic correction terms in the generic class. However as we are working at large N limits we can 
readily see that the second term of the above equation for $\epsilon_2$ dies down  rapidly and we can ignore it's effects 
compared to the first term of $1/N$ at leading order.
Note that the benchmark can be easily retrieved by choosing the parameter $p = 2$ and
keeping leading order contributions of $1/N$.  Executing similar techniques as discussed in previous sections we obtain 
$\kappa$ by equating the above parameters with that of the benchmark model as
\beq
\label{l1}
\kappa=\frac{b}{2}\frac{N_*^p}{\bar N_*^2 \sqrt{\ln{\bar N_*}}\ln^q{N_*}}
\eeq
and also the following relationship 
\beq
\label{l2}
\frac{2}{\bar N_*}=\frac{p}{N}-\frac{q}{N_*\ln{N_*}}.
\eeq
The field excursion in this context comes out to be
 \beq
 \label{delphil}
 \Delta\phi= \sqrt{2}\kappa\int \frac{\ln^{q/2} N}{N^{p/2}}dN
 \eeq

For specific choices of $p$ and $q$ one can infer the implications of the above expression. In the large N limit the second 
slow roll parameter is given by
 \beq
 \label{le21}
\epsilon_2=-\frac{p}{N}.
\eeq

Keeping $p$ fixed we vary the variable $q$ and see how the inflationary field excursion changes. It is to be noted, 
from the various models conforming to the logarithmic class of models and from working within our approximation of 
neglecting the second term of Eqn (\ref{le2}), that only values of 
 $q$ running less than 10 are physically acceptable. The value of $p$ is chosen as 2 which is not only the case for the 
 KM II model but also well motivated from the literature \cite{rb, alpha1, alpha2}. An intensive study of the variation in the 
 inflationary field range with changing $q$ shows that the field excursion remains sub Planckian for parameter range  
 chosen above. Therefore considering all the results obtained in this and in the previous sections what we can conclude 
is that the degeneracy in various pictures may be lifted from future observations aiming at more finer value of tensor-to-scalar ratio 
$r$ also an independent detection of $\epsilon_1$ will help. \\
 
 \noindent
One may wonder why the first Hubble flow function has only been chosen to specify the end of inflation. 
Note that the large $N$ formalism and subsequently the Hubble flow functions considered here are based on the 
primary quantity $H(\phi)$.  The dynamics has been used to define the slow roll parameters instead of the field potentials. 
One can show that the first two Hubble flow functions are linked with corresponding potential slow roll parameters via 
the relations $\epsilon_1 = \epsilon_V$ and $\epsilon_2 = - 4\epsilon_V + 2 \eta_V$.  Liddle $et. al$ in \cite{hsr} have 
pointed out that the true end point of inflation gauged by the Hubble flow functions occur exactly at $\epsilon_1 = 1$ . 
For potential slow roll parameters this is a first order approximation. Now the type of models encompassed by large $N$ formalism 
\cite{rb} exhibit such a dynamics that generically one can assume the end inflation by $\epsilon_1 = 1$.  This
is also the case for models of inflation which consider the existence of flat directions. On the other hand if one still 
gets interested to explore the possibility of ending inflation via alternative methods, one may look for the possibility $\epsilon_2=1$ (
note that this is not same as setting $\eta_V = 1$). This possibility gives rise to a decreasing $\Delta \phi$ w.r.t. $p$ 
in logarithmic class for a given value of $q$. It may be a topic of interest to explore in future endeavors. In those cases one may also 
go beyond the regime of slow roll approximation and look for alternatives like ending inflation by introducing a second field potential.
  
 \section{Summary and Discussions}
 
 Let us conclude with a few comments. We have emphasized on Planck 2015 data and the strong underlying theoretical background 
 in choosing the benchmark model for our analysis. Considering the span of inflaton field profile $\Delta\phi$ for the KM II
 model as reference, we have studied how the range of the field excursion varies in different universality classes of inflationary 
 models corresponding to a chosen point in the $n_s$ -- $r$ plane. The value for $n_s$ satisfies the 
value found by different experiments and also the most recent values given by Planck 2015 \cite{Pl}. The value for the 
tensor-to-scalar ratio $r$ for our benchmark model is well within the allowed upper bound for the value of $r$ as found in 
recent experiments (unlike the chaotic inflation benchmark case). Thus our choice of the chosen point in the $n_s$ -- $r$ plane
 is quite well motivated and future experiments which probe with greater accuracy the value of $r$ can comment on it's viability.
 At present it is in excellent agreement with the experimental results.  The technique followed in this work has been 
 proposed in the context of chaotic inflation as the benchmark model \cite{belli}. However the recent Planck data release rules 
 out values of $r > 0.09$ while for quadratic potential in the chaotic class $r = 0.16$.   
 KM II model predicts a spectral index $n_s$ well within the $2 \sigma$ 
 contour of Planck. This also predicts a value of $r$ that gives sub-Planckian field excursion according to the Lyth bound. 
 
 Comparing 
 this with other universality classes of models we found that it is possible to have super-Planckian as well as 
 sub-Planckian field excursion, for example, for different ranges of parameter $p$ in the perturbative class of models. 
 While equating the slow roll parameters at horizon crossing one not only changes  $\phi_*$ but also  $\phi_{f}$ which 
 are the values of field at the horizon crossing and at the end of inflation respectively.  This also changes the value of $N_{f}$, 
 the number of e-folds at the end of inflation.  Fixing the Hubble flow parameters at horizon crossing for a model 
 amounts  to fixing the value of the field variable $\phi_*$ which in turn changes $N_*$ the number of e-folds at horizon crossing.

 Basically the demand to get the same $n_s$ and $r$ as the benchmark model puts a constraint on the theory via
 the change of $N_*$ and $\phi_*$ from there original value. This is why we get a range of values for $N$ and 
 $\Delta\phi$ for the same $n_s$ and $r$. One can also go ahead and constrain the value of the running of the 
 spectral tilt $\alpha_s$ for the benchmark model and the various classes of inflation. We have checked that to infer that it 
 doesn't introduce any significant constraint for the inflationary field range. For the perturbative and the logarithmic class the third 
 slow roll parameter has a similar $1/N$ dependence as the second slow roll parameter for the two classes of inflation and therefore adds 
 nothing new to the discussion. For the non- perturbative class the third 
 slow roll parameter comes out to be zero. 
 
 In this analysis with KM II as benchmark, most interestingly, we have found that one can get sub-Planckian field excursion 
 in the regime of single field slow roll inflation.  
 There appears to be a maximum value for the field excursion variable and the number of e-folds.  Owing to the different geometric form 
 of the potential in the benchmark model we get a distinct limit on the above mentioned parameters. The chaotic model is 
 much steeper so the rolling down velocity of the field is greater than that for the KM II whose slope is 
 much more flatter leading to a much slower rolling speed. Thus in this case there are more number of 
 e-folds but a smaller value of the maximum field excursion. Interestingly similar results have been observed 
 in non-perturbative class like those in the perturbative class.  Finally we see that for the perturbative class the 
 value of $\Delta\phi_{max}$ is almost same for different initial parameters 
 for the benchmark models while the maximum no. of e-folds changes appreciably. The degeneracy we have observed 
 in different forms may be lifted by future observations \cite{Creminelli}.

 \section{Acknowledgement}

 Argha Banerjee acknowledges the resources provided by Presidency University where a major part of this work was performed. The 
 DST-FIST grant aided library in the Department of Physics at Presidency University has been particularly used for this work.

\end{document}